\documentstyle[11pt]{article}

\title{\bf Topological Vacuum of the Closed Universe as a Gauging Factor}

\author{L. B. Savchenko$^*$, V. A. Savchenko$^{\dag}$ and G. M. Vereshkov \\ {\it Department 
of theoretical physics, Rostov State University,}
\\{\it Zorge str. 5, Rostov-on-Don 344090, Russia}}

\vspace{1.0in}

\date{}

\begin{document}

\maketitle

\begin{abstract}
The only known general base to eliminate the vacuum divergencies of 
quantized matter fields in quantum geometrodynamics is the fermion-boson 
supersymmetry. The topological effect of the closed Universe -- discretization 
of the vacuum fluctuations spectra -- allows to formulate the conditions of  cancellation of the  
divergencies. In the center of attention of this work is the fact that these conditions 
result in the considerable restrictions on the gauge and factor-ordering ambiguities 
peculiar to the equations of the theory and, in the limits of the isotropic model of the 
Universe, remove these ambiguities completely.
\end{abstract}

\vspace{3in}

\noindent $^*$e-mail:savchenko@aaanet.ru

\noindent $^{\dag}$e-mail: savchenko@phys.rnd.runnet.ru

\vfill\eject
\section{Introduction}\label{intro}	
\vspace*{-0.5pt}
\noindent
Quantization of physical fields in a gravitational field of the Universe encounters 
the standard problem of vacuum divergencies. As, however, in quantum 
geometrodynamics (QGD) the standard perturbation theory with its renormalization 
procedure is inapplicable, the only known base for solution of 
the problem is supersymmetry \cite{SS} (its urgency for QGD at present 
is a commonplace). The vacuum contributions of fermion and boson fields 
are opposite on a sign, and the topological effect of a closed universe -- 
discretization of the vacuum fluctuations spectra -- allows to formulate conditions of cancellation 
of the vacuum divergencies. The most remarkable thing is that these conditions sharply 
limit gauge arbitrariness in QGD, in particular, the homogeneous isotropic model turns out
completely deprived of this arbitrariness by them. Moreover, if, taking into account 
the known ambiguity in ordering and parametrization \cite{V,HP} of the 
Wheeler -- DeWitt (WDW) theory \cite{DeWitt}, one imposes on it the mentioned 
conditions, the Hamiltonian of this theory will gain a nontrivial spectrum of eigenvalues, in 
essence conterminous with an energy spectrum of matter including its vacuum 
component. This radically changes the structure of the theory, approximating it to the 
{\it dynamic} one proposed in the work \cite{SSV}. Below, for distinguishing the two 
versions of geometrodynamics, we understand by QGD the dynamic 
theory \cite{SSV}, and the WDW structure, in which there is no time as a formal 
parameter of dynamics, we shall name, as it is accepted now,  the quantum cosmology  
(QC WDW).

In the following section the statement of the problem in the simplest, semiclassical, 
version of a closed homogeneous isotropic universe is adduced. In Sec.~3. 
we shall formulate the problem of topological vacuum spectrum in the closed isotropic 
Universe,  give the outcomes of its solution and show that the effect of topological 
vacuum (TV) can not be coordinated with the semiclassical theory. 
Sections~4. and 5. are devoted to solution of the problem within the 
framework of quantum theory: QC WDW (Sec.~4.) and QGD 
(Sec.~5.).

\textheight=7.8truein
\section{Semiclassical aspect}
\label{semiclass}
\noindent

In the self-consistent semiclassical theory the $ \left \{0\atop 0\right\} $-Einstein equation of a closed 
homogeneous isotropic universe at the presence of matter has the form:
\begin{equation}
\label{semicl1}
\frac 3\kappa \left( \frac{\dot a^2}{{\cal N}^2a^2}+\frac 1{a^2}\right)=\frac
{{\cal N}}{2\pi ^2a^5}\left\langle H_{mat}\right\rangle,  
\end{equation}
where $ \kappa $ is the gravitational constant, $ {\cal N} = \sqrt {g _ {00}} $ is the ``laps 
function'', $ a $ is the scale factor, $ \left\langle H_{mat}\right\rangle $ is the quantum 
average Hamiltonian of material fields, as which we shall consider 
massless conformal two-component objects: neutral vector, 
spinor and charged scalar fields:
\[
H^{\left( s\right) }=\sum_{{\bf k}}H_{{\bf k}}^{\left( s\right) },
\]
$s$ beeng the spin quantum number, \[{\bf k}=\left( \lambda ,j,m\right)
,\;\lambda =1+s,2+s,\ldots ,\infty ,\;j=s,s+1,\ldots ,\lambda-1,\;m=-j,\ldots ,+j.\]

The Hamiltonian of boson fields can be presented as
\[
H_{{\bf k}}^{\left( 0\right) },H_{{\bf k}}^{\left( 1\right) }=\lambda _{{\bf %
k}}\left( a_{{\bf k}}^{\dagger }a_{{\bf k}}+\bar a_{{\bf k}}^{\dagger }\bar
a_{{\bf k}}+1\right), 
\]
where the two annihilation operators $ a _ {{\bf k}}, \, \bar a _ {{\bf k}} $ concern two 
different components of a field (accordingly -- the creation operators  $ a _ {{\bf 
k}} ^ {\dagger}, \, \bar a _ {{\bf k}} ^ {\dagger} $).

The Hamiltonian of fermions
\[
H_{{\bf k}}^{\left( 1/2\right) }=\lambda _{{\bf k}}\left( b_{{\bf k}%
}^{\dagger }b_{{\bf k}}+\bar b_{{\bf k}}^{\dagger }\bar b_{{\bf k}}-1\right). 
\]
 In a state with definite energy
\begin{equation}
\left\langle H^{\left( s\right) }\right\rangle ={\cal E}^{\left( s\right)
}=\sum_{{\bf k}}{\cal E}_{{\bf k}}^{\left( s\right) },\;{\cal E}_{{\bf k}%
}^{\left( s\right) }=\lambda _{{\bf k}}^{\left( s\right) }\left( n_{{\bf k}%
1}^{\left( s\right) }+n_{{\bf k}2}^{\left( s\right) }\pm 1\right), 
\label{energy}
\end{equation}
where ``$ \pm $'', correspondingly, for bosons and fermions, $n_{\bf k}$ are the occupation 
numbers. According to  (\ref {energy}), the quantity $ {\cal \ E} ^ {\left (s\right)} $ 
consists of two physically different parts,
\[
{\cal E}^{\left( s\right) }={\cal E}_{part}^{\left( s\right) }+{\cal E}%
_{vac}^{\left( s\right) },\;
\]
\[
{\cal E}_{part}^{\left( s\right) }=\sum_{{\bf k}}\lambda _{{\bf k}}^{\left(
s\right) }\left( n_{{\bf k}1}^{\left( s\right) }+n_{{\bf k}2}^{\left(
s\right) }\right) 
\]
is the particle energy,
\begin{equation}
{\cal E}_{vac}^{\left( s\right) }=\pm \sum_{{\bf k}}\lambda _{{\bf k}%
}^{\left( s\right) }=\pm \sum_{\lambda =1+s}^\infty \sum_{j=s}^{\lambda
-1}\sum_{m=-j}^j\lambda =\pm \sum_{\lambda =1+s}^\infty \lambda \left(
\lambda ^2-s^2\right)   
\label{vacuum}
\end{equation}
is the vacuum component. The equation (\ref{semicl1}) takes the form:
\[
\frac 3\kappa \left( \frac{\dot a^2}{{\cal N}^2a^2}+\frac 1{a^2}\right) =\frac
{{\cal N}}{2\pi ^2a^5}{\cal E},
\]
${\cal E}=\sum_{s}\left({\cal E}_{part}^{(s)}+{\cal E}_{vac}^{(s)}\right)$. In 
particular, in the gauge  of ``conformal time'' 
\begin{equation}
\label{gauge}
{\cal N}=a
\end{equation}
we have
\begin{equation}
\label{semicl}
\frac{6\pi ^2}\kappa \left( \dot a^2+a^2\right) ={\cal E}.
\end{equation}

As was mentioned in Introduction, the diverging vacuum contributions (\ref {vacuum}) 
of various fields can not be renormalized by the standard quantum field methods.
However opposition of signs of these contributions from bosons and fermions specifies 
that the cancellation of divergencies is possible in the framework of the boson-fermion 
supersymmetry \cite{SS}.

\section{Topological vacuum. Inconsistency of semiclassical approximation}
\label{TV}

\noindent
Let us consider an elementary supermultiplet of the two-component boson field and the 
two-component fermion one. According to (\ref{vacuum}), the vacuum energy of $N$ 
scalar-spinor supermultiplets is determined by the
\begin{equation}
\label{sum1}
{\cal E}_{vac}=\sum_{a=1}^{2N}\sum_{\lambda =1}^\infty \lambda
^3-\sum_{a=1}^{2N}\sum_{\lambda =3/2}^\infty \lambda \left( \lambda ^2-\frac
14\right) .
\end{equation}

To reduce the sums to common limits let us make in the spinor sum the 
replacement $ \lambda \rightarrow \lambda + 1/2 $, and then in both sums $ \lambda ^ 
3$ put shifts $ \lambda = k + \alpha _ a $ with $ \alpha _ a $ integers:
\[
\sum_a\sum_{\lambda =1}^\infty \lambda ^3=\sum_a\sum_{k=1}^\infty \left(
k+\alpha _a \right)^3. 
\]
The shifts $ \alpha _ a $ should be chosen so that the divergencies were mutually
compensated, and then the residual $ {\cal E} _ {vac} $ will develop out of the sums
$ \sum _ {k = 1-\alpha _ a} ^ 0 $ if $ \alpha _ a> 0 $, and $ -\sum _ {k = 1} ^ {-\alpha 
_ a} $ if $ \alpha _ a < 0 $. The degrees of the divergencies are determined by the formulas:
\[
\sum_{k=1}^Kk^3=\frac 14K^2(K+1)^2;
\]
\[
\sum_{k=1}^Kk^2=\frac 16K\left( K+1\right) \left( 2K+1\right) ;
\]
\[
\sum_{k=1}^Kk=\frac 12K\left( K+1\right) .
\]

So, 
\[
\frac 16AK\left( K+1\right) \left( 2K+1\right) +\frac 12BK\left( K+1\right)
+CK=0;
\]
\[
A=3\sum_{a=1}^{N}\left( \alpha _a-\beta _a\right) -3\frac N2,
\]
\[
B=3\sum_{a=1}^{N}\left( \alpha _a^2-\beta _a^2\right) -\frac N2,
\]
\[
C=\sum_{a=1}^{N}\left( \alpha _a^3-\beta _a^3\right) ,
\]
$\alpha _a$ are the shifts in the scalar sums, $\beta _a$ are those in the
spinor sums. 

The condition of divergency cancellation is $ A = B = C = 0 $, that means the 
set of equations:
\[
\sum_{a=1}^{N}\left( \alpha _a-\beta _a\right) =\frac N2,
\]
\[
\sum_{a=1}^{N}\left( \alpha _a^2-\beta _a^2\right) =\frac N6,
\]
\[
\sum_{a=1}^{N}\left( \alpha _a^3-\beta _a^3\right) =0.
\]

The second version of divergency cancellation appears after the replacement 
$ \lambda \to \lambda-1/2 $ in the 
spinor sum (\ref {sum1}). In this case the set of equations 
for $ \alpha_ a, \beta _ a $ takes the form
\[
\sum_{a=1}^{N}\left( \alpha _a-\beta _a\right) =-\frac N2,
\]
\[
\sum_{a=1}^{N}\left( \alpha _a^2-\beta _a^2\right) =\frac N6,
\]
\[
\sum_{a=1}^{N}\left( \alpha _a^3-\beta _a^3\right) =0.
\]

For $ N $ vector-spinor supermultiplets there are 4 different versions in
correspondence with the two possible appropriate replacements of $ \lambda $ in each of 
the two sums -- boson and fermion:
\begin{enumerate}
\item 
\[
\sum_{a=1}^{N}\left( \alpha _a-\beta _a\right) =-\frac N2,
\]
\[
\sum_{a=1}^{N}\left( \alpha _a^2-\beta _a^2\right) =-\frac N2,
\]
\[
\sum_{a=1}^{N}\left( \alpha _a^3-\beta _a^3\right) =0.
\]

\item 
\[
\sum_{a=1}^{N}\left( \alpha _a-\beta _a\right) =-3\frac N2,
\]
\[
\sum_{a=1}^{N}\left( \alpha _a^2-\beta _a^2\right) =-\frac N2,
\]
\[
\sum_{a=1}^{N}\left( \alpha _a^3-\beta _a^3\right) =0.
\]

\item 
\[
\sum_{a=1}^{N}\left( \alpha _a-\beta _a\right) =-\frac N2,
\]
\[
\sum_{a=1}^{N}\left( \alpha _a^2-\beta _a^2\right) =\frac N2,
\]
\[
\sum_{a=1}^{N}\left( \alpha _a^3-\beta _a^3\right) =0.
\]

\item 
\[
\sum_{a=1}^{N}\left( \alpha _a-\beta _a\right) =\frac N2,
\]
\[
\sum_{a=1}^{N}\left( \alpha _a^2-\beta _a^2\right) =\frac N2,
\]
\[
\sum_{a=1}^{N}\left( \alpha _a^3-\beta _a^3\right) =0.
\] 
\end{enumerate}
 
These sets of equations are solvable at $ N $ multiple 12, and the vector-spinor sets 1. 
and 2. are solvable at $ N $ multiple 8 as well. The vacuum spectrum $ {\cal E} _ {vac} $ obtained 
as a result of solution of these sets of equations for various $ N $ has proved to be {\it 
equidistant}, extending from $ -\infty $ up to $ + \infty $ with the same pitch equal 6:
\begin{equation}
\label{vac}
{\cal E}_{vac}=6n+v,\;n=0,\pm 1,\pm 2,\ldots, 
\end{equation}
and $ v $ can take values $ 0, \pm 2 $ depending on a set of supermultiples.

Coming back to the semiclassical equation (\ref {semicl}), we must note the 
following. The  $ {\cal E} _ {part} $ being a function of the occupation numbers, the $ {\cal 
E}_{vac}$ is an independent variable, and the negative half of its rigid 
discrete spectrum cannot be coordinated with the positive definite left part 
of the equation. So, the semiclassical approach to the physics of the early 
Universe is incomplete: the presence of the TV demands appealing to the full quantum 
theory.
 
\section{The Wheeler -- DeWitt quantum cosmology.}
\label{WDW}

\noindent

The WDW equation
\begin{equation}
\label{WDWeq}
H\Psi=0
\end{equation}
can be arranged in such a way that the operator $H$ divides into tree parts,
\[
H=H_{grav}+H_{part}+{\cal E}_{vac},
\]
where $H_{grav}$ is its gravitational component,
$H_{part}=\sum_i\sum_{\bf k}H_{\bf k}^{(i)},
H_{\bf k}^{(i)}=\lambda_{\bf k}^{(i)}a_{\bf k}^{i\dag}a_{\bf k}^i$,
the index $i$ numbers components of material fields, $a _ {\bf k} ^ i$ are the 
annihilation operators of the appropriate particles. The solution of the 
equation (\ref {WDWeq}) can be factored: 
$ \Psi = \Psi _ {part} \cdot\Psi _ {grav} $, so that
\begin{equation}
\label{part}
H_{part}\Psi_{part}={\cal E}_{part}\Psi_{part},
\end{equation}
\begin{equation}
\label{grav}
H_{grav}\Psi_{grav}=({\cal E}_{grav}-{\cal E}_{part}-{\cal E}_{vac})\Psi_{grav}.
\end{equation}

Thus, an eigenvalue $ {\cal E} _ {part} $ of the operator $ H _ {part} $ has to be 
considered as a parameter of the equation (\ref{grav}). As to the spectrum 
$ {\cal E} _ {vac} $, it should be obtained as a result of solution of the WDW equation
responsible for a state of the Universe as a whole. Fortunately, parametrization/gauge 
and permutation ambiguities of this equation allow to come near the solution of the 
problem.

At first, the oscillator aspect of the left part of the equation (\ref {semicl}) and 
the spectrum nature of the$ {\cal E} _ {vac} $ (\ref{vac}) specify the definite 
advantage of the gauge (\ref {gauge}). If, in addition, one puts $ {\cal E} _ {part} = 0$ 
and chooses a system of units in which $6\pi^2/\kappa=1/2$, 
the appropriate WDW equation can be presented as:
\[
-\frac 12\left( \frac{\partial ^2}{\partial a^2}-a^2\right) \Psi_{vac}= 
E_{vac}\Psi _{vac},\;a\geq 0, 
\]
\[
E_{vac}=2m+3/2,\;m=0,1,2,\ldots.
\]
Comparing the spectrum of eigenvalues of this equation with (\ref {vac}), we
should, first of all, remove all the values of the $m$ that are not multiple 3. 
Existing $ {\cal E}_{part} \neq 0 $ requires presence at the left part of the 
equation of an element that could absorb this component. And here comes to 
the aid the second ambiguity of the 
quantum theory in general and of the WDW equation in particular, known as the 
ordering problem. Here it manifests itself in  that in the general case the kinetic part
of the WDW Hamiltonian can be presented as
\[
\frac12f^{-1}\left( a\right) \frac
\partial {\partial a}f^2\left( a\right) \frac \partial {\partial
a}f^{-1}\left( a\right),
\]
where $ f (a) $ is an arbitrary differentiable function of the scale factor.
Let us put $ f = a ^ {l + 1} $. Then the WDW equation takes the form of the equation
for a ``centrifugal oscillator''
\[
\left[ \frac 12\left( \frac{\partial ^2}{\partial a^2}-a^2-\frac{%
l\left( l+1\right) }{a^2}\right) +{\cal E}\right] \Psi _{{\cal E}%
}=0
\]
with the spectrum (comp. \cite{LL})
\begin{equation}
\label{spec}
{\cal E}=2m+l+\frac32,\;l={\cal E}_{part}.
\end{equation}

What do we have in the result? Presence of the TV with its distinctive spectrum (\ref 
{vac})
\begin {enumerate}
\item fixes the gauge and solves the ordering problem;
\item transforms QC WDW to the theory with a nontrivial spectrum.
\end {enumerate}
Thus, the TV has properties intrinsic to the gravitational vacuum condensate (GVC) 
which existence the consecutive construction of QGD on the basis of the usual quantum 
theory \cite{SSV} inevitably leads to.

On the other hand, the WDW equation cuts the vacuum spectrum removing,
in particular,
the negative values $ n $, but does not contain any absorption mechanism for 
this part of the energy spectrum of the subsystem ``topological vacuum''. This 
is a superfluous testimony of the stated in \cite{SSV} incompleteness of the 
theory.

The other situation is in QGD.

\section{Quantum geometrodynamics}
\label{QGD}
\noindent

In the dynamic Schr\"odinger equation for the physical part of a wave function in
the gauge (\ref{gauge}) the Hamiltonian has the form:
\begin{equation}
H_{ph}=\frac 18\frac 1{M\left( a\right) }\frac
\partial {\partial a}M\left( a\right) \frac \partial {\partial a}-2a^2,
\label{Hamilt}
\end{equation}
$M$ being a probability measure. Let us put
\begin{equation}
M=a^{2\left( l+1\right)},l\geq 0.  
\label{measure}
\end{equation}
The probability measure, beeng as well a measure of a path integral (PI), is generated by 
the certain dependence  on $ a $ of a pitch $ \epsilon $ of splitting up the time interval 
at a final PI definition through a multiple integral, as the measure is connected with $ 
\epsilon $ by the relation \cite{SSV}, which in this case has the form
\[
\gamma^{-\frac 12}M={\rm const}, \gamma=a^{4(l+1)},
\]
where $\gamma$ is defined by
\[
\epsilon =\chi\gamma,\;\chi \rightarrow 0.
\]

By substituting the measure (\ref {measure}) into the Hamiltonian (\ref {Hamilt}), we 
come to the equation for stationary states
\[
\left[ \frac 18\frac{\partial ^2}{\partial a^2}+\frac 1{4a}\frac \partial
{\partial a}-2a^2-\frac{l\left( l+1\right) }{8a^2}+{\cal E}%
\right] \Psi=E\Psi,
\]
$E$ being a parameter of GVC \cite{SSV}. The eigenvalue spectrum of the equation
is defined by the same expression (\ref{spec}),
\[
{\cal E}-E={\cal E}_{part}+{\cal E}_{vac}-E=2n+l+\frac32,\;n=0,1,2,\ldots;\;\;l={\cal E}_{part},
\]
but this time there is the GVC parameter $E$ allowing to put it in agreement 
with the TV:
presenting (\ref{vac}) in the form ${\cal E}_{vac}=6(n-m+v),\;n,m=1,2,\ldots$ 
we have
\begin{equation}
\label{spectrum}
E=4n-6m+6v-\frac32.
\end{equation}
Thus, GVC and 
TV represent a united subsystem, breaking gauge symmetry concerning the 
diffeomorphism group. 

\section{Conclusion.}
\noindent

The above-stated analysis shows that presence of material fields with physical topological vacuum 
breaks the 
basic equation of the WDW theory -- the quantum version of the classical Hamilton 
constraint, -- revealing GVC. This confirms our conclusion made in the 
work \cite{SSV} that the 
quantum principle in cosmology is incompatible to the principle of gauge invariance 
concerning the diffeomorphism group and that QC WDW has not any 
general-theoretical foundation.

Here we have considered the elementary cosmological model, which, however,
most likely can serve a good ``zero approximation'' to the exact model
of the early Universe. It definitely allows to expect that the obtained
results, though will not be so hard for the real model, nevertheless, in
essential will be applicable to it. Or else, it is possible to state, that
TV considerably limits the gauge/parametrization arbitrariness and the ordering 
ambiguity (choice of a probability measure) in QGD.

\end{document}